\documentclass[reprint,onecolumn,aps,
showpacs,notitlepage]{revtex4-1}

\usepackage[utf8x]{inputenc}
\usepackage[T1]{fontenc}
\usepackage{mathptmx} 
\usepackage{bigints}
\usepackage{caption}
\usepackage{subcaption}
\usepackage{psfrag}
\usepackage{grffile}
\usepackage{verbatim}
\usepackage{microtype}
\usepackage{multirow}
\usepackage{enumitem}
\usepackage{amsmath}
\usepackage{amssymb}
\usepackage{amsthm}
\usepackage{bbm}
\usepackage{dsfont}
\usepackage{mathtools}
\usepackage{graphicx}

\usepackage[linesnumbered,ruled]{algorithm2e}
\usepackage{color}
\usepackage{lineno}
\usepackage{xcolor}
\usepackage{soul}
\usepackage{enumitem} 

\definecolor{myblue}{rgb}{0.153,0.322,0.706}
\usepackage[colorlinks,linkcolor=myblue,urlcolor=myblue,citecolor=myblue]{hyperref}
\usepackage{geometry}
\newcommand{\be}{\begin{equation}}
\newcommand{\ee}{\end{equation}}

\newcommand{\fra}[1]{{\color{pink} Francesco: #1}}

\hypersetup{ 	
pdfsubject = {},
pdftitle = {},
pdfauthor = {}
}

\begin{document} 

\title{Delocalization-localization dynamical phase transition of random walks on graphs}

\author{Giorgio Carugno}
\email{giorgio.carugno@kcl.ac.uk}
\affiliation{Department of Mathematics, King’s College London, Strand, London WC2R 2LS, UK}

\author{Pierpaolo Vivo}
\email{pierpaolo.vivo@kcl.ac.uk }
\affiliation{Department of Mathematics, King’s College London, Strand, London WC2R 2LS, UK}

\author{Francesco Coghi}
\email{francesco.coghi@su.se}
\affiliation{Nordita, KTH Royal Institute of Technology and Stockholm University, Hannes Alfvéns väg 12, SE-106 91 Stockholm, Sweden}

\date{\today}


\begin{abstract}
We consider random walks evolving on two models of connected and undirected graphs and study the exact large deviations of a local dynamical observable. We prove, in the thermodynamic limit, that this observable undergoes a first-order dynamical phase transition (DPT). This is interpreted as a `co-existence' of paths in the fluctuations that visit the highly connected bulk of the graph (delocalization) and paths that visit the boundary (localization). The methods we used also allow us to characterize analytically the scaling function that describes the finite size crossover between the localized and delocalized regimes.  Remarkably, we also show that the DPT is robust with respect to a change in the graph topology, which only plays a role in the crossover regime. All results support the view that a first-order DPT may also appear in random walks on infinite-size random graphs.
\end{abstract}

\maketitle

\section{Introduction}

Random walks on graphs are versatile tools to model real-world noisy dynamical processes embedded in spatial structures \cite{Hughes1995,Noh2004,Barrat2008,Newman2010,Latora2017,Masuda2017}. These processes describe both natural and man-made phenomena such as the spreading of infectious diseases \cite{Barrat2008,Pastor-Satorras2015}, the transport of vesicles in cell cytoskeletons \cite{Julicher1997}, the propagation of information in communication networks \cite{Castellano2009,Liu2014}, and the robustness of networks to random failures \cite{Barrat2008} to name just a few examples. Often, the focus in these applications is towards time-averaged quantities including stationary distributions, and energy and particle currents. Indeed, these are observables commonly used in applications to gather information on the average state occupation and mobility in network structures \cite{Barrat2008,Masuda2017}. On the other hand, fluctuations are also fundamental to understand the behavior of physical systems living in unstable environments as rare events are often responsible for the evolution dynamics \cite{Albeverio2006,Kishore2011}. However, much less is known about them and in the last decades many research efforts have been deployed towards the development of theoretical frameworks that allow for their study, e.g., large deviation theory \cite{DenHollander2000,Touchette2009,Dembo2010,Chetrite2015,Jack2020a,Carugno2022}.

Recently, signatures of a dynamical phase transition (DPT), viz.\ a transition between different fluctuation mechanisms, has been identified in the study of the mean degree (connectivity) visited by unbiased random walks evolving on sparse random graphs \cite{DeBacco2016,Coghi2019,Gutierrez2021}. There are good grounds to consider it as a first-order DPT where we observe the coexistence of two `phases' characterized by random walk paths that visit the whole graph, and paths localized in dangling chains, i.e., lowly connected structures of the graph. However, a rigorous proof for ensembles of random graphs is still lacking and, in fact, the community still debates on the real nature and interpretation of DPTs \cite{Whitelam2018,Whitelam2021}.

In this paper, we contribute to the debate by analyzing two exactly solvable models where the transition appears to be first-order and characterized by an absorbing dynamics. This sees, on the one hand, the random walk fully localized in dangling structures, and on the other hand, the random walk fully absorbed by the bulk of the graph, which acts as an entropic basin and allows the random walk to be fully delocalized. We make use of a theoretical framework for the calculation of large deviations that we developed in \cite{Carugno2022} and that allows us to: (i) consider general time-additive observables, (ii) analytically characterize the behavior of random walks on finite-size graphs, and (iii) rigorously study the scaling (with respect to the size of the graph) of fluctuations around the critical value of the DPT. Remarkably, in agreement with \cite{Whitelam2018,Whitelam2021} we notice that an important ingredient for the appearance of a first order DPT in both models is the presence of absorbing dynamics, generated by different scalings of the hopping probabilities in the graph. Furthermore, we notice that although the first order DPT appears in both the models we investigated, the scaling of the fluctuations around the transition is different and we argue that it is both function of the dynamical process and of the inherent topology of the network.

A brief outline of the paper follows. In Section \ref{sec:theory} we set up a general model of an URW hopping on a graph, discuss the general form of observables that we consider in this manuscript, and introduce the theory of large deviations in this setting. In Section \ref{sec:results} we collect our results related to two exactly solvable models. In Section \ref{sec:conclusion} we conclude the paper by summarizing the results obtained and briefly discussing open questions.

\section{Setting and large deviations}
\label{sec:theory}

We consider an unbiased discrete-time random walk (URW) $X = \left( X_{\ell} \right)_{\ell=1}^{n} = (X_1,X_2,\dots,X_{n})$ evolving on a finite connected graph $G = (V,E)$, with $V$ denoting the set of $N$ vertices (or nodes) and $E$ the set of edges (or links). The topology of the graph is encoded in the symmetric adjacency matrix $A$, which has components
\begin{equation}
\label{eq:Adjacency}
A_{ij} = 
 \begin{cases} 
      1 & i \in \partial j \\
      0 & \text{otherwise} \, ,
   \end{cases}
\end{equation}
where $\partial j$ denotes the set of neighbors of node $j$. Notice that we choose to consider an unweighted symmetric graph for simplicity, but our methods can be easily generalized to more structured cases. The dynamics of the random walk is defined by the transition matrix $\Pi$ having components
\begin{equation}
    \label{eq:Trans}
    \Pi_{ij} = \frac{A_{ij}}{k_i} \, ,
\end{equation}
where $k_i = \sum_{j \in V} A_{ij}$ is the degree of node $i$, viz.\ the number of edges in which node $i$ participates. The matrix $\Pi$ characterizes the uniform probability of going from a vertex $X_{\ell}=i$ at time $\ell$ to a vertex $X_{\ell+1}=j$ at time $\ell+1$ -- that is, the probability of transitioning from $i$ to $j \in \partial i$ does not depend on $j$. Furthermore, for simplicity we restrict the random walk to be ergodic, viz.\ $\Pi$ is irreducible and aperiodic. In the rest of the manuscript we use the index $\ell$ to refer to time and the indices $i$ and $j$ to refer to nodes of the graph.

The long-time behavior of the URW is well understood. Thanks to ergodicity, the random walk has a unique stationary distribution
\begin{equation}
    \label{eq:StatDistr}
    \rho_i = \frac{k_i}{\sum_{j \in V} k_j} \, ,
\end{equation}
which is found to be proportional to the degree of each node. Furthermore, the URW is also reversible, viz.\ it is an equilibrium process, as it satisfies the detailed balance condition
\begin{equation}
    \label{eq:Detailed}
    \rho_i \Pi_{ij} = \rho_j \Pi_{ji} \, 
\end{equation}
for each pair of nodes in $V$.

In this setting, we assume that the URW $X$ accumulates a cost in time given by
\begin{equation}
\label{eq:Cost}
C_n = \frac{1}{n} \sum_{\ell=1}^n f(X_{\ell}) \, ,
\end{equation}
where $f$ is any function of the vertex state. In nonequilibrium statistical mechanics, this cost is also called a dynamical observable \cite{Touchette2009} and, depending on $f$, it may represent interesting physical quantities, such as occupation times \cite{Chetrite2015}, internal energy \cite{Sekimoto2010}, chemical concentrations \cite{Dykman1998}, activities \cite{Gutierrez2021}, and entropy production rates \cite{Coghi2019}.
Because of the ergodicity of the URW, in the long-time limit the observable $C_n$ converges with probability $1$ to the ergodic average
\begin{equation}
    \label{eq:ErgodicAv}
    \sum_{i \in V} \rho_i f(i) \eqqcolon c^* \, .
\end{equation}
This convergence property is often used to estimate properties of large graphs such as degree distributions or centrality measures, by running random walks (or, generally speaking, agents) on the graphs for long times \cite{Newman2010}.

Following the introduction, here we study fluctuations of $C_n$ around the typical value $c^*$ by calculating its probability distribution $\mathbb{P}(C_n=c)$ in the $n \rightarrow \infty$ limit. The probabilistic theory of large deviations \cite{DenHollander2000,Touchette2009,Dembo2010} tells us that this distribution has an exponentially decaying form
\begin{equation}
    \label{eq:ExponentiallyDecaying}
    \mathbb{P}(C_n=c) = e^{-n I(c) + o(n)} \, ,
\end{equation}
described by the rate function $I$ given by the following limit
\begin{equation}
    \label{eq:Rate}
    I(c) = - \lim_{n \rightarrow \infty} \frac{1}{n} \log \mathbb{P}(C_n=c) \, .
\end{equation}
The rate function $I$ is a pivotal object in the theory of large deviations as it characterizes the fluctuations of $C_n$ to leading order in $n$; it is a non-negative function and it is equal to $0$ for ergodic random walks only at $c^*$ (where the probability concentrates exponentially fast with time).

Much effort is drawn towards the development of methods that allow one to calculate $I$ in \eqref{eq:Rate} efficiently \cite{Touchette2009}. Spectral and variational techniques can both be implemented and, depending on the particular model studied, it may well be that some work better than others \cite{Coghi2021PhD}. Spectral techniques based on moment generating functions have the merit to reformulate the problem in a different setting---similarly to a microcanonical-canonical change of ensemble---whereas variational techniques based on the contraction principle \cite{DenHollander2000,Touchette2009} are useful to find probabilistic bounds \cite{Hoppenau2016}. In the following, we will base our large deviation study on the techniques discussed in \cite{Carugno2022} which try to merge the pros of spectral and variational methods.

In line with \cite{Carugno2022} and previous works \cite{Whittle1955,Dawson1957,Goodman1958,Billingsley1961,CsiszaR1987,DenHollander2000,Polettini2015,Dembo2010}, in order to calculate the rate function $I$ associated with the observable $C_n$ in \eqref{eq:Cost} we move the focus on to the study of the higher-dimensional pair-empirical occupation measure
\begin{equation}
    \label{eq:PairEmp}
    L^{(2)}_n(i,j) = \frac{1}{n} \sum_{\ell=1}^n \delta_{X_{\ell},i}\delta_{X_{\ell+1},j}  = \nu_{ij} \hspace{1cm} \forall i,j \in V \, ,
\end{equation}
which counts the fraction of jumps $\nu_{ij}$ that the URW makes between each couple of nodes in the graph -- see \cite{Carugno2022}. Remarkably, the value of $C_n$ can be deduced via the formula
\begin{equation}
\label{eq:ContrForm}
C_n = \sum_{i,j \in V} f(i) L_n^{(2)}(i,j) \, .
\end{equation}
We can calculate the rate function \eqref{eq:Rate} by means of the G\"{a}rtner--Ellis theorem \cite{Touchette2009, Dembo2010}. To do so we need to introduce the scaled cumulant generating function (SCGF) of $C_n$, which is defined as
\begin{equation}
    \label{eq:SCGFDef}
    \lambda_{s,N}[\nu^*] = \lim_{n \rightarrow \infty} \frac{1}{n} \mathbb{E} \left[ e^{n s C_n} \right] \ ,
\end{equation}
and calculate its Legendre--Fenchel transform, i.e.,
\begin{equation}
\label{eq:LegendreFenchel}
I(c) = \sup_{s \in \mathbb{R}} \{ sc - \lambda_{s,N}[\nu^*] \} \, .
\end{equation}
In \cite{Carugno2022}, we showed that $\lambda_{s,N}[\nu^*]$ can be obtained minimizing the following action
\begin{eqnarray}
    \label{eq:FinAct}
    &\lambda_{s,N} &[\nu] = \lambda_{1,N}[\nu]+ \lambda_{2,N}[\nu]+ \lambda_{3,N}[\nu]+ \lambda_{4,N}[\nu]\\
    \label{eq:l1}
    &\lambda_{1,N}&[\nu] = \sum_{i=1}^N\sum_{j=1}^N \nu_{ij} \left( \log\left(\sum_{k=1}^N \nu_{ik}\right) - \log(\nu_{ij}) \right) \\
    \label{eq:l2}
    &\lambda_{2,N}&[\nu] = \sum_{i=1}^N \sum_{j=1}^N \log (\Pi_{ij}) \  \nu_{ij}\\
    \label{eq:l3}
    &\lambda_{3,N}&[\nu] = s \sum_{i=1}^N f(i) \sum_{j=1}^N \ \nu_{ij}\\
    \label{eq:l4}
    &\lambda_{4,N}&[\nu] = \epsilon \left( \sum_{i=1}^N \sum_{j=1}^N \nu_{ij} - 1 \right)   + \sum_{i=1}^N \eta_i \left(\sum_{j=1}^N \nu_{ij}- \sum_{j=1}^N \nu_{ji}\right) \ ,
\end{eqnarray}
with respect to $\nu_{ij}$, $\epsilon$ and $\eta_i$, which are respectively the fraction of jumps from node $i$ to node $j$, and the Lagrange multipliers fixing the normalization constraint and the global balance. We remark that these formulas are valid for any finite $N$.
In our setting the rate function $I$ in \eqref{eq:LegendreFenchel} reduces to
\begin{equation}
    \label{eq:RateAct}
    I(c) = - \lambda_{1,N}[\nu^*] - \lambda_{2,N}[\nu^*] - \lambda_{4,N}[\nu^*] \, ,
\end{equation}
where the dependence on the fluctuation $c$ enters through the minimizer $\nu^*$, which depends on the optimized tilting parameter $s^*(c)$, i.e., $\nu^* \equiv \nu^*(s^*(c))$. For further details on the methods we used to derive \eqref{eq:FinAct}-\eqref{eq:l4} and on a useful physical characterization of the action \eqref{eq:FinAct} we refer the reader to \cite{Carugno2022} and related bibliography.

Although the equations for the minimum of \eqref{eq:FinAct} may be complicated to solve analytically for complex models, the proposed approach has several advantages that will also be highlighted in the next sections when studying simplified scenarios. Firstly, the explicit form of the action \eqref{eq:FinAct} for any finite $N$ allows us to study directly the scaling of the fluctuations with varying graph size. As we will see in the following, this is an important feature that will help us in characterizing fluctuations around critical points. Furthermore, the tilting parameter $s$ is responsible for biasing the dynamics of the URW via \eqref{eq:l3} to realize a fluctuation $c$ for the observable $C_n$ fixed by the Legendre duality equation
\begin{equation}
    \label{eq:LegendreDuality}
    c = \frac{d \lambda_{s,N}[\nu]}{d s} \, .
\end{equation}
Therefore, the minimizer $\nu^*$ of the action $\lambda_{s,N}$ characterizes the typical configuration of jumps on the graph $G$ that give rise to the fluctuation $c$ defined by \eqref{eq:LegendreDuality} (similarly to \cite{Gutierrez2021}). \\
It is natural to introduce a biased dynamics for which $C_n = c$ is realized in the typical state: this biased dynamics has been thoroughly characterized in its general form in \cite{Chetrite2015,Chetrite2015a} and also in the setting of random walks on graphs in \cite{Coghi2019}. The process characterized by the biased dynamics is known as driven (or effective/auxiliary) process and in this context is a locally-biased version of the URW whose transition probability matrix has been described in \cite{Coghi2019}. Within our approach, we can also fully define the driven process and its transition matrix, which reads
\begin{equation}
    \label{eq:DrivenTransition}
    \left( \Pi_s \right)_{ij} = \frac{\nu^*_{ij}(s)}{\sum_{k \in V} \nu^*_{ik}(s)} \, .
\end{equation}
In other words, the driven process is the effective biased random walk that explains how a fluctuation $C_n=c$ is created up to time $n$.

We remark that the method used here to calculate the SCGF $\lambda_{s,N}[\nu^*]$ via the minimization of \eqref{eq:FinAct} is equivalent to spectral methods based on the so-called tilted matrix \cite{Touchette2009,Dembo2010,Touchette2018,Coghi2021PhD}. In particular, in \cite{Carugno2022} we show that the Euler--Lagrange equations for the minima of \eqref{eq:FinAct} are a useful re-writing of the dominant eigenvalue problem associated with the tilted matrix. The main pros of the method reviewed in \cite{Carugno2022} are (i) to give a clear physical interpretation of all the terms in the action and SCGF and (ii) to express the driven process in terms of the minimizers of \eqref{eq:FinAct}, which are the optimal jumps that create a fluctuation $C_n=c$.

\section{Delocalization-localization dynamical phase transition}
\label{sec:results}

Recently, it has been pointed out \cite{DeBacco2016,Coghi2019,Gutierrez2021} that an URW that accumulates a cost proportional to the degree of each visited node, e.g., $f(X_l) = k_{X_l}$ in \eqref{eq:Cost}, and that runs on the largest connected component of an Erd\H{o}s--R\'{e}nyi random graph seems to undergo a DPT. This transition is localized in the fluctuations of the mean degree visited when this is lower than the mean connectivity of the graph and is interpreted as a `co-existence' of paths that visit nodes with low degree and paths that visit the whole graph \cite{DeBacco2016,Coghi2019}. [Noticeably, another DPT may arise when the random walk visits more often highly connected regions of the graph and localizes around the highest degree node of the graph. Although as interesting, in this manuscript we will not focus on this behavior.] The Erd\H{o}s--R\'{e}nyi random graph in question is picked from a `canonical' ensemble of random graphs having a fixed number of nodes $N$ and edges randomly placed with a small probability $p$ between each pair of nodes such that $N p = \bar{c}$ is fixed to be the mean degree of the graph. Noticeably, signatures of the DPT disappear when $\bar{c}$ is large, revealing that a fundamental topological ingredient for the appearance of the transition is the presence of lowly connected structures in the graph (such as trees and dangling chains of nodes). These structures carry strong spatial correlations---a node of degree one is likely to be connected with a node of degree two in a dangling chain and these correlations are responsible for the dynamics of the random walk when visiting low-degree nodes. The overall picture is that of a random walk whose behavior fluctuates between two distinct phases characterized by (i) being localized in lowly-connected regions of the graph and (ii) being spread over the bulk (most connected region) of the graph which acts as an ergodic basin absorbing the dynamics (see model in Appendix A of \cite{Coghi2019} and also \cite{Whitelam2018}).

As far as the current state of the art is concerned, it is not clear whether such a DPT appears in the infinite-size limit of ensemble of random graphs. However, as mentioned in the previous paragraph, various numerical studies indicate an abrupt change in the mechanisms that generate fluctuations, endorsing the idea of a dynamical phase transition \cite{DeBacco2016,Coghi2019,Gutierrez2021}.

In the following, by applying the theory discussed in Section \ref{sec:theory}, we analytically characterize the DPT in two models, which catch what we think are the most relevant physical features of this phenomenon. We believe that these characteristics are shared by the dynamics of URWs on Erd\H{o}s--R\'{e}nyi random graphs: in particular, the heterogeneity of the scaling of the degree and the presence of lowly connected regions such as dangling chains. We show that (i) the DPT is first order -- that is, $\lambda_{s,N}[\nu^*]$ as a function of $s$ has a non-differentiable point $s_c$ in which the first derivative is discontinuous; (ii) the behavior around $s_c$ is characterized by a scaling function which is not universal, and depends on both the dynamics of the URW and the topology of the graph; (iii) the driven process can be fully characterized, allowing us to understand how fluctuations arise in time. These results give further evidence of the presence of a first-order DPT in random walks exploring random graphs. 

\subsection{Bulk-dangling model}

\begin{figure}
    \centering
    \includegraphics[width=0.4\linewidth]{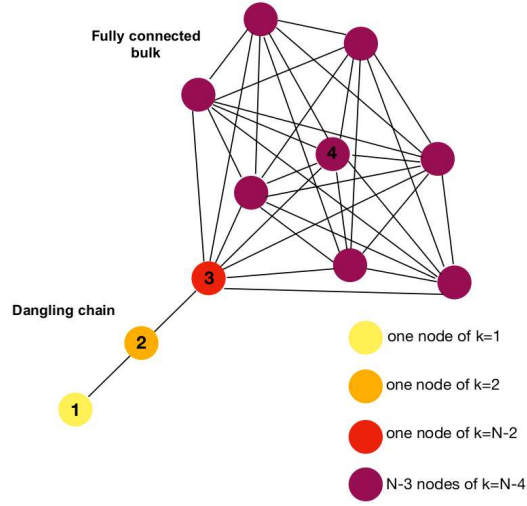}
    \caption{Bulk-dangling model graph for $N = 11$. Node $1$ has degree $k_1 = 1$, node $2$ has degree $k_2 = 2$ and node $3$ has degree $k_3 = N-2 = 9$: the first two nodes represent the dangling chain, while node $3$ bridges the chain with the bulk, being part of the latter. All the remaining $N-3 = 8$ nodes are of type $4$, having degree $k_4 = N - 3 = 8$. Together with node $3$, they form the fully connected bulk.}
    \label{fig:BulkDanglingGraph}
\end{figure}

The first model we look at is based on an URW with transition matrix \eqref{eq:Trans} collecting a cost \eqref{eq:Cost} of the form
\begin{equation}
    \label{eq:BulkDanglingCost}
    C_n = \frac{1}{n} \sum_{\ell=1}^n \frac{k_{X_\ell}}{\bar{k}} \, ,
\end{equation}
by visiting a graph of $N$ nodes composed by a fully connected bulk of $N-2$ nodes and a single dangling chain of $2$ nodes (see Fig.\ \ref{fig:BulkDanglingGraph}). The structure of this graph incorporates two relevant features: (i) the presence of a spatially correlated dangling chain and (ii) a fully connected bulk that allows the URW to uniformly spread over the network. Given this structure, the mean degree of the graph is $\bar{k} = ((N-3)(N-3)+(N-2) + 2 + 1)/N$, which evidently scales linearly with $N$ for large-size graphs. This feature allows us to deduce the behavior of the observable $C_n$ in two opposite scenarios in the $N \rightarrow \infty$ limit. Evidently, if the random walk is uniformly spread over the bulk $C_n \sim 1$, whereas if the random walk is localized in the chain $C_n \sim 0$. We argue that this behavior does not depend on the length $L$ of the dangling chain -- the choice $L = 2$ is made to simplify calculations.

Following Section \ref{sec:theory} we calculate the action $\lambda_{s,N}[\nu]$ in \eqref{eq:FinAct}. Because of the bulk of the graph being highly symmetric, i.e., every link in the bulk is equivalent, and the global balance imposed on the dynamics, i.e., incoming and outgoing flow of a node being equal, we are only left with four degrees of freedom (variables) $\nu_{ij}$  that determine the action. We name the fraction of jumps
\begin{eqnarray}
    \label{eq:BulkDangling1}
    &\nu_{12} & \;\; \text{for both directions: $1 \rightarrow 2$ and $2 \rightarrow 1$} \\
    \label{eq:BulkDangling2}
    &\nu_{23} & \;\; \text{for both directions: $2 \rightarrow 3$ and $3 \rightarrow 2$}\\
    \label{eq:BulkDangling3}
    &\nu_{34} & \;\; \text{for both directions of each link: $3 \rightarrow 4$ and $4 \rightarrow 3$} \\
    \label{eq:BulkDangling4}
    &\nu_{44} & \;\; \text{for both directions of each link in the bulk} \, .
\end{eqnarray}
Notice that if at this stage we also imposed the normalization constraint, i.e., $\sum_{i,j \in V} \nu_{ij} = 1$, we would be left with only three degrees of freedom. However, for practical reasons in the calculation of the minimum of the action $\lambda_{s,N}[\nu]$, we leave this last constraint as an implicit parametrization with a Lagrange multiplier $\epsilon$ entering the action.

The action can be explicitly written as
\begin{equation}
    \label{eq:BulkDanglingAction}
    \lambda_{s,N}[\nu] = h(s,N,\nu_{12},\nu_{23},\nu_{34},\nu_{44}) + \epsilon (1 - 2 \nu_{12} - 2 \nu_{23} - 2 (N-3) \nu_{34} - (N-3)(N-4) \nu_{44}) \, ,
\end{equation}
with the long form of the function $h$ postponed to the Appendix \ref{ref:app}. The minimum and minimizers of the action \eqref{eq:BulkDanglingAction} can be found by solving the saddle-point equations that can be cast in the following linear system:
\begin{align}
\label{eq:BulkDanglingSystem1}
& \nu_{12} =  \nu_{23} \frac{a(s,N,\epsilon)}{1-a(s,N,\epsilon)} \\
\label{eq:BulkDanglingSystem2}
& \nu_{23} =  \nu_{34} \frac{b(s,N,\epsilon)}{1-b(s,N,\epsilon)} \\
\label{eq:BulkDanglingSystem3}
& \nu_{34} =  \nu_{23} \frac{c(s,N,\epsilon)}{1-(N-3)c(s,N,\epsilon)} \\
\label{eq:BulkDanglingSystem4}
& \nu_{44} =  \nu_{34} \frac{d(s,N,\epsilon)}{1-(N-4)d(s,N,\epsilon)} \\
\label{eq:BulkDanglingNormConstraint}
& 2\nu_{12} + 2\nu_{23} + 2(N-3) \nu_{34} +  (N-3)(N-4) \nu_{44} = 1 \, ,
\end{align}
with
\begin{align}
&a(s,N,\epsilon) = e^{3 \frac{s}{\bar{k}}-\log 2 +2 \epsilon} \\
&b(s,N,\epsilon) = \frac{e^{N \frac{s}{\bar{k}} -\log 2 - \log(N-2) +2\epsilon} }{ 1 - e^{3 \frac{s}{\bar{k}} - \log 2 +2 \epsilon} }\\
&c(s,N,\epsilon) = \frac{e^{(2N-5) \frac{s}{\bar{k}} - \log(N-2) - \log(N-3) +2\epsilon}}{1 - e^{(N-3) \frac{s}{\bar{k}} - \log ( N-3) + \epsilon} } \\
&d(s,N,\epsilon) = e^{(N-3) \frac{s}{\bar{k}} - \log ( N-3) + \epsilon} \, .
\end{align}
We report the form of the minimizers $\nu^*=(\nu^*_{12},\nu^*_{23},\nu^*_{34},\nu^*_{44})$ in the Appendix \ref{ref:app}. It is important to remark that these minimizers are not yet in a fully explicit form as they depend on the Lagrange parameter $\epsilon$ (which, in \cite{Carugno2022}, has also been proved to be the negative SCGF $\lambda_{s,N}[\nu^*]$). However, the Lagrange parameter $\epsilon$---hence, also the SCGF we are after---can be determined by solving the normalization constraint in \eqref{eq:BulkDanglingNormConstraint} after having replaced the form of the minimizers $\nu^*=(\nu^*_{12},\nu^*_{23},\nu^*_{34},\nu^*_{44})$. The equation reads
\begin{equation}
    \label{eq:BulkDanglingDeterminant}
    \begin{split}
    2(N-2)(N-3) - 2 &\tau (N-2)(N-4) e^{(N-3)\frac{s}{\bar{k}}} - \tau^2 (N-3) \left((N-2)e^{3 \frac{s}{\bar{k}}} + e^{N \frac{s}{\bar{k}}} + 2 e^{(2N-5) \frac{s}{\bar{k}}} \right) +\\
    & \tau^3 (N-4) \left((N-2)e^{N \frac{s}{\bar{k}}} + e^{(2N-3) \frac{s}{\bar{k}}} \right) + \tau^4 (N-3) e^{(2N - 2) \frac{s}{\bar{k}}} = 0 \, ,
\end{split}
\end{equation}
with $\tau = e^{\epsilon}$. Noticeably, \eqref{eq:BulkDanglingDeterminant} can also be derived by imposing that the matrix of the coefficients of the four linear equations \eqref{eq:BulkDanglingSystem1}--\eqref{eq:BulkDanglingSystem4} has a nullspace of dimension greater than zero---that is, when its determinant is zero. Equation \eqref{eq:BulkDanglingDeterminant} is fourth order in $\tau$ and hence it admits four solutions of which only one is physical. This can be selected by noticing that, since $\nu^*$ must be positive, the right hand side of the four equations \eqref{eq:BulkDanglingSystem1}--\eqref{eq:BulkDanglingSystem4} is also positive (we postpone the exact form of the inequality constraints to the Appendix \ref{ref:app}). In this way, we obtained the SCGF $\lambda_{s,N}[\nu^*]$ valid for any finite-size graph.

\begin{figure}[tb]
\captionsetup[subfigure]{justification=centering}
\begin{subfigure}{.33\textwidth}
  \centering
  \psfrag{a}{$s$}
  \psfrag{b}[][Bl][1][180]{$\lambda_{s,N}$}
  \psfrag{c}{$N=15$}
  \psfrag{d}{$N=25$}
  \psfrag{e}{$N=100$}
  \psfrag{f}{$N \rightarrow \infty$}
  \includegraphics[width=0.95\linewidth]{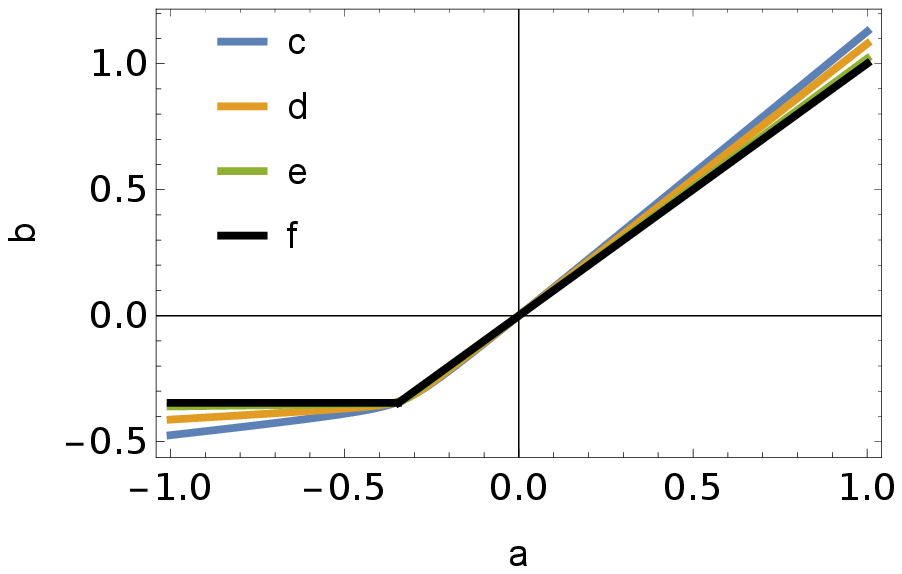}
  
  \caption{SCGF $\lambda_{s,N}$}
  \label{fig:BulkDanglingSCGF}
\end{subfigure}
\begin{subfigure}{.33\textwidth}
  \centering
  \psfrag{a}{$s$}
  \psfrag{b}[][Bl][1][180]{$\lambda'_{s,N}$}
  \includegraphics[width=0.95\linewidth]{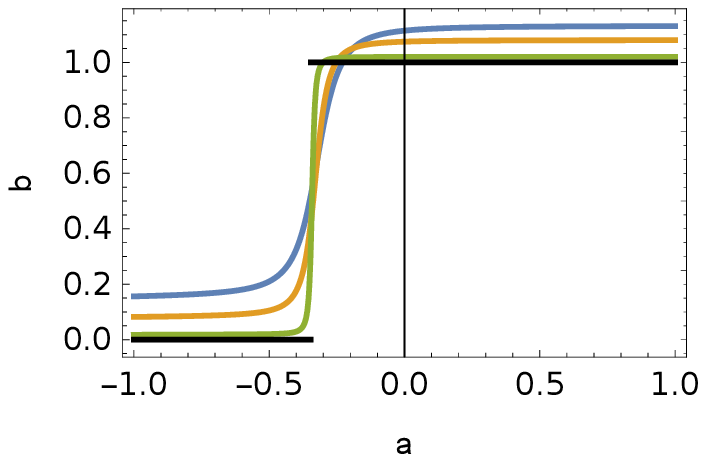}  
  \caption{Derivative of SCGF $\lambda'_{s,N}$}
  \label{fig:BulkDanglingSCGFDer}
\end{subfigure}
\begin{subfigure}{.33\textwidth}
  \centering
  \psfrag{a}{$c$}
  \psfrag{b}[][Bl][1][180]{$I(c)$}
  \includegraphics[width=0.95\linewidth]{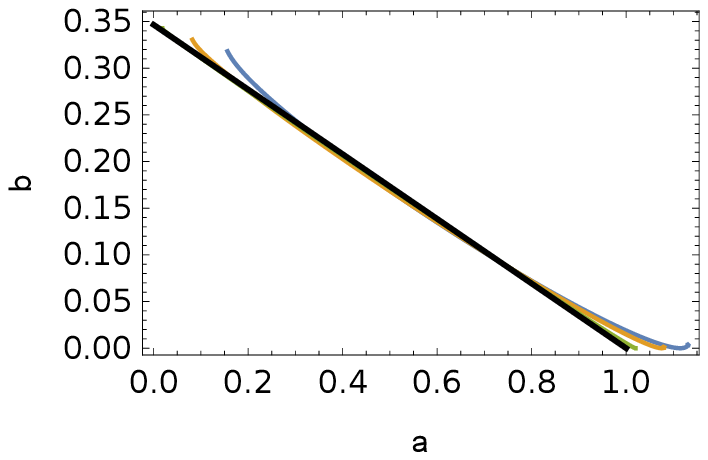}  
  \caption{Rate function $I(c)$}
  \label{fig:BulkDanglingRate}
\end{subfigure}
\caption{Large deviation study for the bulk-dangling model. In all three figures, different colors correspond to a different number $N$ of nodes: i) light blue is $N=15$; ii) orange is $N = 25$; iii) green is $N = 100$; iv) black is $N \to \infty$. All finite $N$ curves where obtained by solving \eqref{eq:BulkDanglingDeterminant} numerically, while analytical expressions for the black curves are presented in \eqref{eq:BulkDanglingSCGFInfinite} for figure (a), \eqref{eq:BulkDanglingSCGFInfiniteDer} for figure (b) and \eqref{eq:BulkDanglingRateInfinity} for figure (c). }
\label{fig:BulkDangling}
\end{figure}
By carefully taking the limit $N \rightarrow \infty$ in the polynomial equation \eqref{eq:BulkDanglingDeterminant} we can also explicitly obtain the SCGF in the infinite-size limit of the graph, which is
\begin{equation}
    \label{eq:BulkDanglingSCGFInfinite}
    \lambda_{s,\infty} = 
    \begin{cases} 
      - \frac{\log 2}{2} & s \leq -\frac{\log 2}{2} \\
      s & s > -\frac{\log 2}{2} \, 
   \end{cases}
\end{equation}
and highlights a non-differentiable point at $s_c = - \log 2/2$. The derivative of $\lambda_{s,\infty}$, according to \eqref{eq:LegendreDuality}, is
\begin{equation}
    \label{eq:BulkDanglingSCGFInfiniteDer}
    \frac{d \lambda_{s,\infty}}{d s} = 
    \begin{cases} 
      0 & s \leq -\frac{\log 2}{2} \\
      1 & s > -\frac{\log 2}{2} \, ,
   \end{cases}
\end{equation}
which explicitly describes the fluctuation $C_n = c$ happening with varying tilting parameter $s$. This confirms our expectations: on the left of the critical point $s_c$ the random walk is localized in the dangling chain---the only region of the graph where the cost accumulated $C_n$ (see \eqref{eq:Cost}) does not scale with the size $N$---whereas on the right of $s_c$ the random walk is spread in the bulk where it accumulates a cost that scales linearly with $N$. Furthermore, the value of $s_c$---and with it all the left branch of $\lambda_{s,\infty}$ in \eqref{eq:BulkDanglingSCGFInfinite}---can easily be interpreted as the mean entropy of the random walk that, localized in the dangling chain, keeps going back and forth from the node of degree $1$ to the node of degree $2$ (see also \cite{Carugno2022}). Eventually, the rate function $I$ can easily be obtained by Legendre transforming the two analytical branches of \eqref{eq:BulkDanglingSCGFInfinite} and by connecting them with a linear section or by implementing directly \eqref{eq:Rate}; in both cases we obtain
\begin{equation}
    \label{eq:BulkDanglingRateInfinity}
    I(c) = 
    \begin{cases} 
      \frac{\log 2}{2} - c \frac{\log 2}{2} & 0 \leq c \leq 1 \\
      \infty & \text{otherwise} \, ,
   \end{cases}
\end{equation}
and we remark that the non-differentiable point $s_c$ for the SCGF $\lambda_{s,\infty}$ is mapped onto the linear section characterizing the rate function. We graphically show in Fig.\ \ref{fig:BulkDangling} the SCGF, its derivative, and the rate function for finite-size graphs and in the infinite-size limit.

The non-differentiability of the SCGF can be physically related to a first-order DPT that is interpreted here as a coexistence of paths that either visit predominantly the bulk of the graph ($C_n \sim 1$) or are localized in the dangling chain ($C_n \sim 0$). A further characterization of this DPT is given by identifying the mechanisms that give rise to the fluctuations around the critical point $s_c$.

As it appears from formula \eqref{eq:BulkDanglingSCGFInfiniteDer} and Fig.\ \ref{fig:BulkDanglingSCGFDer}, the normalized mean-degree visited by the URW \eqref{eq:BulkDanglingCost} computed from \eqref{eq:LegendreDuality} is a piece-wise constant function of the tilting parameter $s$ in the infinite-size limit of the graph. We refer to the region $s < s_c$ ($s>s_c$) corresponding to the localized (delocalized) phase as $s^-$ ($s^+$) and we study in these two regions the scaling with $N$ of the transition probabilities of the driven process \eqref{eq:DrivenTransition}. The calculation can be done by properly taking the $N \rightarrow \infty$ limit of the minimizer $\nu^* = (\nu^*_{12},\nu^*_{23},\nu^*_{34},\nu^*_{44})$ and inserting the result in \eqref{eq:DrivenTransition}. We get the following two transition matrices that characterize the probability of stepping from a node to another one in the graph of Fig.\ \ref{fig:BulkDanglingGraph}:
\begin{equation}
\label{eq:BulkDanglingDriven1}
\Pi_{s^-} = \left( \begin{array}{cccccc} 0 & 1 & 0 & \cdots & 0 & \cdots \\ 1 + O(N^{-1}) & 0 & O(N^{-1}) & \cdots & 0 & \cdots \\ 0 & -\frac{e^s}{3 s} + O(N^{-1}) & 0 & O(N^{-1}) & \cdots & \cdots \\ \vdots & & \ddots & & \ddots & \cdots \\ 0 & 0 & 1 -\sqrt{2}e^s & 0 & O(N^{-1}) & \cdots \end{array} \right)
\end{equation}
\begin{equation}
\label{eq:BulkDanglingDriven2}
\Pi_{s^+} = \left( \begin{array}{cccccc} 0 & 1 & 0 & \cdots & 0 & \cdots \\ \frac{e^{-2s}}{2} + O(N^{-1}) & 0 & \left( 1 - \frac{e^{-2s}}{2} \right) + O(N^{-1}) & \cdots & 0 & \cdots \\ 0 & O(N^{-1}) & 0 & O(N^{-1}) & \cdots & \cdots \\ \vdots & & \ddots & & \ddots & \cdots \\ 0 & 0 & O(N^{-1}) & 0 & O(N^{-1}) & \cdots \end{array} \right) \, .
\end{equation}
Evidently, for fluctuations obtained by fixing $s < s_c$ the random walk is biased towards localizing in the dangling chain, e.g., if the random walk is on node two, the probability of hopping onto node one is one order of magnitude (with respect to the system size) bigger than moving towards the fully connected bulk. For $s>s_c$ instead, the bulk behaves as an entropic basin absorbing the random walk and allowing it to be fully spread over the graph.
\begin{figure}
    \centering
    \psfrag{a}{$\tilde{s}$}
    \psfrag{b}[][Bl][1][180]{$\tilde{\epsilon}_{\tilde{s}}$}
    \psfrag{c}{$N=15$}
    \psfrag{d}{$N=25$}
    \psfrag{e}{$N=100$}
    \psfrag{f}{$N \rightarrow \infty$}
    \includegraphics[width=0.5\linewidth]{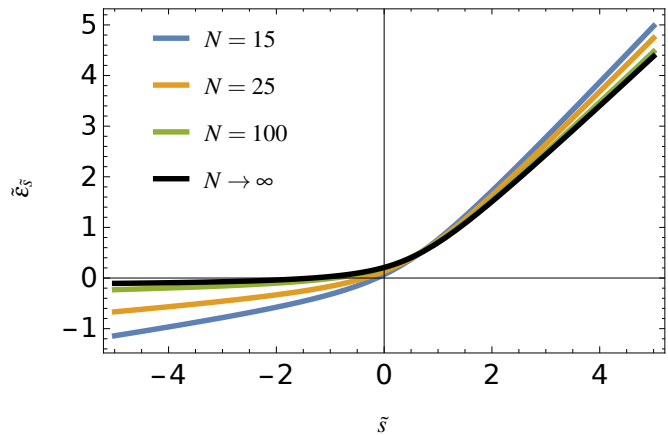}
    \caption{Crossover regime of the SCGF of the bulk-dangling model around $s_c$ as a function of the scaling variable $\tilde{s}$ for different values of $N$. As $N$ increases, the colored curves collapse into the limiting curve predicted theoretically \eqref{eq:BulkDanglingCorrection}.
    }
    \label{fig:BulkDanglingZoom}
\end{figure}

We conclude the study of the bulk-dangling model showing how fluctuations scale with the graph size locally around the critical point $s_c$ (in analogy with the study carried out in \cite{Whitelam2018}). This can be done by centering and rescaling the tilting variable $s$ as
\begin{equation}
    \label{eq:BulkDanglingScalings}
    s = -\frac{\log 2}{2} + \frac{\tilde{s}}{N} \, ,
\end{equation}
and the Lagrange parameter (or negative SCGF)
\begin{equation}
    \label{eq:BulkDanglingScalingSCGF}
    \epsilon = \frac{\log 2}{2} - \frac{\tilde{\epsilon}_{\tilde{s}}}{N} \, ,
\end{equation}
in the polynomial equation \eqref{eq:BulkDanglingDeterminant}. Using this scaling and expanding the polynomial to leading order in $N$ we obtain

\begin{equation}
    \label{eq:BulkDanglingSCGFLocal}
    \lambda_{\tilde{s},N} \approx - \frac{\log 2}{2} + \frac{\tilde{\epsilon}_{\tilde{s}}}{N} \, ,
\end{equation}
with 
\begin{equation}
    \label{eq:BulkDanglingCorrection}
    \tilde{\epsilon}_{\tilde{s}} = \frac{1}{8} \left( \sqrt{2} + 4 \tilde{s} - 7 \log 2 + \sqrt{2 + 16 \sqrt{2} + 2 \sqrt{2} \log 2 - \log^2 2 - 8 \sqrt{2} \tilde{s} - 8 \log 2 \tilde{s} + 16 \tilde{s}^2} \right) \, ,
\end{equation}
which explains how the SCGF locally scales as a function of the graph size around $s_c$. We report in Fig.\ \ref{fig:BulkDanglingZoom} the function $\tilde{\epsilon}_{\tilde{s}}$ (translated to be centered in $(0,0)$ and not in $(- \log 2 / 2, - \log 2 / 2$). Evidently, the function continuously joins the two branches of fluctuations separated by the critical point $s_c$ in Fig.\ \ref{fig:BulkDanglingSCGF}: on the left, for $\tilde{s} \ll 0$, $\tilde{\epsilon}_{\tilde{s}}$ tends to $0$ (hence, $- \log 2 / 2$)), on the right, for $\tilde{s} \gg 0$, it behaves linearly with respect to $\tilde{s}$.

In conclusion, the critical point $s_c$ marks a first-order DPT for the observable $C_n$ in \eqref{eq:BulkDanglingCost}, however, thanks to a proper rescaling showed in \eqref{eq:BulkDanglingScalings} and \eqref{eq:BulkDanglingScalingSCGF} we can get more precise information on how fluctuations scale with the system size around the critical point $s_c$.

\subsection{Two-state Markov chain}

\begin{figure}
    \centering
    \includegraphics[width=0.6\linewidth]{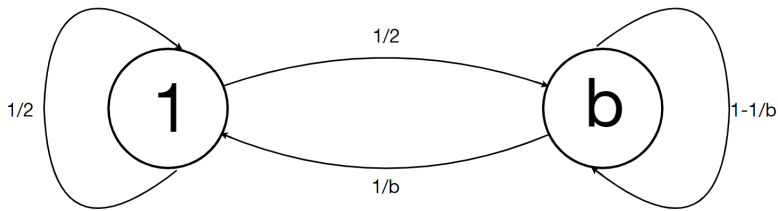}
    \caption{Sketch  of the two-state Markov chain model, composed by state $1$ and state $b$. The transition probabilities -- depicted in the figure above the arrows -- read explicitly: $p_{11} = 1/2$, $p_{1b} = 1/2$, $p_{b1} = 1/b$, $p_{bb} = 1 - 1/b$.}
    \label{fig:2StateModel}
\end{figure}

In this Subsection we analyze another model which takes its cue from the findings in the previous model and is also inspired by the works of \cite{Whitelam2018,Coghi2019}. The model is a two-state Markov chain as represented in Fig.\ \ref{fig:2StateModel}. If the Markov chain is found on the state on the left, namely $1$, at a certain time $\ell$, it collects a unitary reward $1$, whereas if it is on the right, namely $b$, it collects a reward $b \geq 1$ (eventually $b \rightarrow \infty$). Further, although the probability of moving from the left to the right is totally unbiased, the probability of moving from the right to the left inversely scales with the reward $b$. 

The observable we focus on has the general form in \eqref{eq:Cost} and in this particular scenario it reduces to
\begin{equation}
    \label{eq:TwoStateCost}
    C_n = \frac{1}{n} \sum_{\ell = 1}^n \frac{X_\ell}{b} \, ,
\end{equation}
which is the mean reward collected over time renormalized by the maximum reward $b$.

This model tries to catch once again the most relevant physical ingredients that may lead to a delocalization-localization first-order DPT. In doing so, however, we take a further simplification: we try to rule out as much as we can the graph topology, replacing bulk and dangling contributions with two single states which respectively give a reward of $b$ and $1$ to the observed cost in \eqref{eq:TwoStateCost}. In comparison with the previous model, the reward $b$ should be analogous to the graph size $N$---a random walk lost in the bulk of a graph observes nodes with a degree that scales with $N$ in \eqref{eq:BulkDanglingCost}---whereas the reward $1$ should mimic the observed degree in the dangling chain. Furthermore, the inverse scaling with $b$ of the probability of moving from the right state to the left one, should give rise to an absorbing dynamics in the right state for $b \rightarrow \infty$. As we will see in the following, the topology of the graph does not play a pivotal role in the appearance of the first-order DPT, but it may play an important role in determining the exact fluctuations in the crossover regime around the critical point.

Once again, following Section \ref{sec:theory} we calculate the action $\lambda_{s,b}[\nu]$ in \eqref{eq:FinAct}. Because the global balance is imposed on the dynamics, we only have to deal with three degrees of freedom (variables) $\nu_{ij}$ that determine the action. These are
\begin{eqnarray}
    \label{eq:TwoState1}
    &\nu_{11} & \;\; \text{fraction of jumps from $1$ to $1$} \\
    \label{eq:TwoState2}
    &\nu_{1b} & \;\; \text{for both directions: $1 \rightarrow b$ and $b \rightarrow 1$} \\
    \label{eq:TwoStateb}
    &\nu_{bb} & \;\; \text{fraction of jumps from $b$ to $b$} \, .
\end{eqnarray}
Notice that if we also imposed the normalization constraint, i.e., $\sum_{i,j \in \left\lbrace 1,b \right\rbrace} \nu_{ij} = 1$, we would be left with only two degrees of freedom. However, analogously to the previous Subsection, we leave this last constraint as an implicit parametrization with a Lagrange multiplier $\epsilon$ entering the action.

The action can be explicitly written as
\begin{equation}
    \label{eq:TwoStateAction}
    \begin{split}
    \lambda_{s,b}[\nu] &= - \nu_{11} \log \nu_{11} - \nu_{bb} \log \nu_{bb} - 2 \nu_{1b} \log \nu_{1b} + (\nu_{1b} + \nu_{bb}) \log (\nu_{1b} + \nu_{bb}) + (\nu_{1b} + \nu_{11}) \log (\nu_{1b} + \nu_{11}) + \\
    &\hspace{1cm} + \frac{s}{b} (\nu_{11} + \nu_{1b}) + s (\nu_{1b} + \nu_{bb}) - \log 2 \nu_{11} + \log \frac{(b - 1)}{b}  \nu_{bb} + \nu_{1b} (- \log 2 - \log b) + \epsilon (2 \nu_{1b} + \nu_{11} + \nu_{bb} - 1) \, .
    \end{split}
\end{equation}
The minimum and minimizers of the action \eqref{eq:TwoStateAction} can be found by solving the saddle-point equations and imposing the normalization constraint. We get
\begin{align}
\label{eq:TwoStateSystem1}
& \nu_{11} =  \frac{e^{\frac{s}{b} + \epsilon} (-b + (b-1) e^{s+\epsilon})}{-4 b + 2 (b -1) e^{s + \epsilon} + b e^{\frac{s}{b} + \epsilon}} \\
\label{eq:TwoStateSystem2}
& \nu_{bb} = \frac{(b-1) e^{s + \epsilon} (-2 + e^{\frac{s}{b} + \epsilon})}{-4 b + 2 (b -1) e^{s + \epsilon} + b e^{\frac{s}{b} + \epsilon}} \\
\label{eq:TwoStateSystem3}
& \nu_{1b} = \frac{1}{\frac{b}{b-(b-1)e^{s + \epsilon}} - \frac{2}{-2 + e^{\frac{s}{b} + \epsilon}}} \, ,
\end{align}
as still functions of the Lagrange multiplier $\epsilon$. This last can be determined by, for instance, using the equation for the minimum of $\lambda_{s,b}[\nu]$ w.r.t $\nu_{1b}$ and by replacing the values of $\nu_{11}$ and $\nu_{bb}$ with those in \eqref{eq:TwoStateSystem1} and \eqref{eq:TwoStateSystem2}. We find that the SCGF $\lambda_{s,b}[\nu^*]$ is analytically given by
\begin{equation}
    \label{eq:TwoStateSCGF}
    \lambda_{s,b} = - \epsilon = \log \left[ \frac{1}{4b} \left( b e^{\frac{s}{b}} + 2 (b-1)e^{s} + \sqrt{4 (b-1)^2 e^{2 s} + b^2 e^{\frac{2s}{b}} - 4 (b-3) b e^{\frac{b s + s}{b}} } \right) \right] \, .
\end{equation}

By carefully taking the limit $b \rightarrow \infty$ of \eqref{eq:TwoStateSCGF} we explicitly obtain the SCGF in the infinite size limit of the reward, which reads
\begin{equation}
    \label{eq:TwoStateSCGFInfinite}
    \lambda_{s,\infty} = 
    \begin{cases} 
      - \log 2 & s \leq -\log 2 \\
      s & s > -\log 2 \,
   \end{cases}
\end{equation}
and highlights the appearance of a non-differentiable point at $s_c = - \log 2$ (this value can always be interpreted as the mean entropy of the random walk localized in $1$). The derivative of $\lambda_{s,\infty}$, as in \eqref{eq:LegendreDuality}, explicitly describes the fluctuation $C_n = c$ happening with varying tilting parameter $s$ and analogously to \eqref{eq:BulkDanglingSCGFInfinite} we obtain
\begin{equation}
    \label{eq:TwoStateSCGFInfiniteDer}
    \frac{d \lambda_{s,\infty}}{d s} = 
    \begin{cases} 
      0 & s \leq -\log 2 \\
      1 & s > -\log 2 \, .
\end{cases}
\end{equation}
This says that on the left of the critical point $s_c$ the Markov chain is localized in the left state where it accumulates a cost that does not scale with the reward $b$, whereas on the right of $s_c$ the Markov chain is localized in the right state where it accumulates a cost $b$ at every step. We remark that at finite $N$ the critical point is absent, replaced by a crossover region where the Markov chain visits both nodes for a finite fraction of time.

\begin{figure}[ht]
\captionsetup[subfigure]{justification=centering}
\begin{subfigure}{.33\textwidth}
  \centering
  \psfrag{a}{$s$}
  \psfrag{b}[][Bl][1][180]{$\lambda_{s,b}$}
  \psfrag{c}{$b=15$}
  \psfrag{d}{$b=50$}
  \psfrag{e}{$b=250$}
  \psfrag{f}{$b \rightarrow \infty$}
  \includegraphics[width=0.95\linewidth]{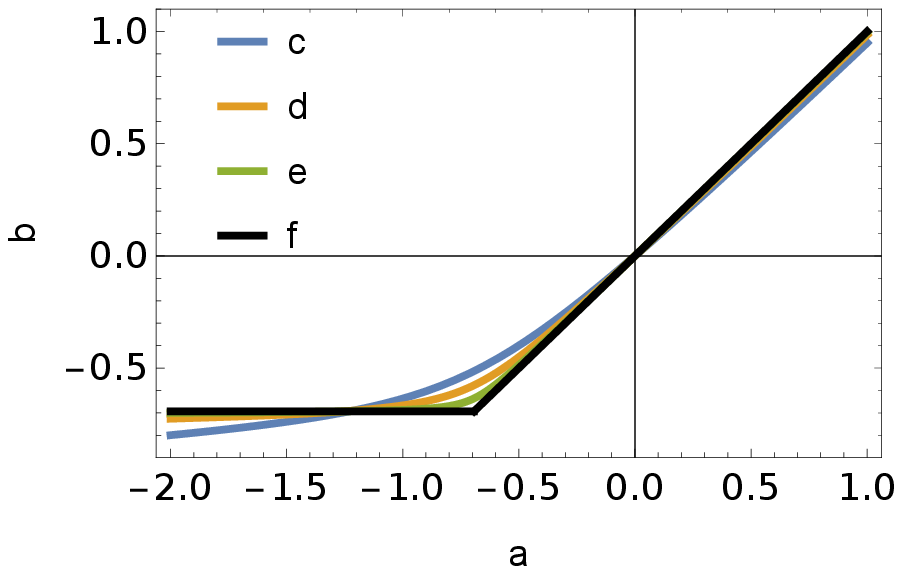}
  \caption{SCGF $\lambda_{s,b}$}
  \label{fig:TwoStateSCGF}
\end{subfigure}
\begin{subfigure}{.33\textwidth}
  \centering
  \psfrag{a}{$s$}
  \psfrag{b}[][Bl][1][180]{$\lambda'_{s,b}$}
  \includegraphics[width=0.95\linewidth]{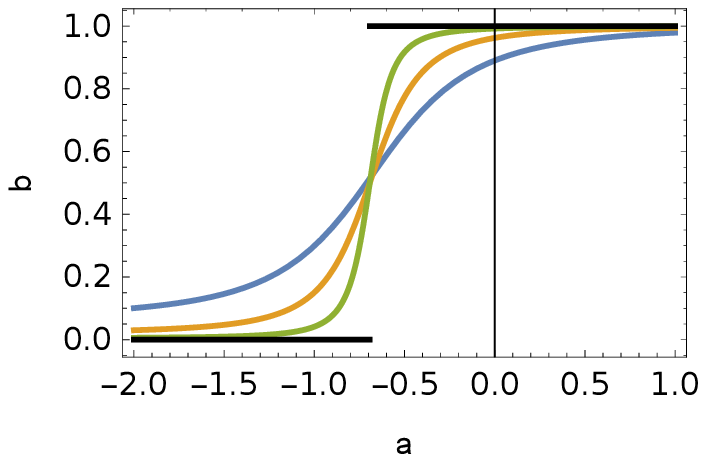}  
  \caption{Derivative of SCGF $\lambda'_{s,b}$}
  \label{fig:TwoStateSCGFDer}
\end{subfigure}
\begin{subfigure}{.33\textwidth}
  \centering
  \psfrag{a}{$c$}
  \psfrag{b}[][Bl][1][180]{$I(c)$}
  \includegraphics[width=0.95\linewidth]{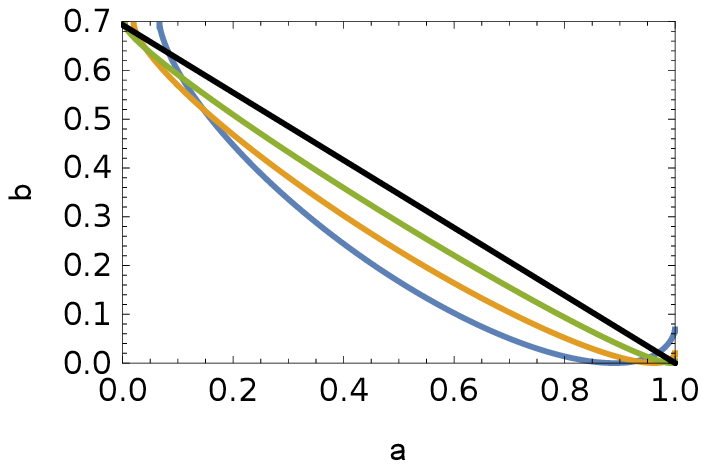}
  \caption{Rate function $I(c)$}
  \label{fig:TwoStateRate}
\end{subfigure}
\caption{Large deviation study for the two-state model. In all three figures, different colors correspond to a different value $b$ of the reward: i) light blue is $b=15$; ii) orange is $b = 50$; iii) green is $b = 250 $; iv) black is $b \to \infty$. All finite $b$ curves where obtained from \eqref{eq:TwoStateSCGF}, while analytical expressions for the black curves are presented in \eqref{eq:TwoStateSCGFInfinite} for figure (a), \eqref{eq:TwoStateSCGFInfiniteDer} for figure (b) and \eqref{eq:TwoStateRateInfinity} for figure (c). }
\label{fig:TwoState}
\end{figure}

The rate function $I$ can also be easily obtained as explained in the previous Subsection and reads
\begin{equation}
    \label{eq:TwoStateRateInfinity}
    I(c) = 
    \begin{cases} 
      \log 2 - c \log 2 & 0 \leq c \leq 1 \\
      \infty & \text{otherwise} \, .
   \end{cases}
\end{equation}
We graphically show in Fig.\ \ref{fig:TwoState} the SCGF, its derivative, and the rate function for the finite reward case and in the infinite-reward $b$ limit. Noticeably, the rate function obtained in \eqref{eq:BulkDanglingRateInfinity} is exactly half the rate function obtained above here. This is consequence of the mean entropy $\lambda_{1,N}+\lambda_{2,N}$ \eqref{eq:l1}, \eqref{eq:l2} that the random walk has in the localized state: in the bulk-dangling model is half with respect to the two-state model presented here.

The non-differentiability of the SCGF can be physically related to a first-order DPT also in this case. Once again, this is interpreted as a coexistence of paths that are either absorbed by the state $b$ ($C_n \sim 1$) or are localized in the state $1$ ($C_n \sim 0$). We can further characterize this DPT by writing the driven process \eqref{eq:DrivenTransition} that leads to fluctuations for $s^- \equiv s < s_c$ or for $s^+ \equiv s > s_c$. This can be done by properly taking the $b \rightarrow \infty$ limit of the minimizer $\nu^* = (\nu^*_{11},\nu^*_{1b},\nu^*_{bb})$ and inserting the results in \eqref{eq:DrivenTransition}. We get the following two transition matrices:
\begin{equation}
\label{eq:TwoStateDriven1}
\Pi_{s^-} = \left( \begin{array}{cc} 1 + O(b^{-1}) & O(b^{-1}) \\ 1-2e^s + O(b^{-1}) & 2e^s + O(b^{-1}) \end{array} \right)
\end{equation}

\begin{equation}
\label{eq:TwoStateDriven2}
\Pi_{s^+} = \left( \begin{array}{cc} \frac{1}{2e^s} + O(b^{-1}) & 1 - \frac{1}{2e^s} + O(b^{-1}) \\ O(b^{-1})  & 1 + O(b^{-1}) \end{array} \right) \, .
\end{equation}
For $s < s_c$ the Markov chain is biased towards localizing in the state $1$, whereas for $s>s_c$, the state $b$ absorbs the Markov chain. This is very similar to what we have seen in the bulk-dangling model, with the only difference that now the role of the topology has been replaced by different rewards on the two states of the chain. 

Although this structural change in the model does not seem to affect the appearance of a first-order DPT, we notice that fluctuations scale differently around the critical point $s_c = - \log 2$. This is made evident by rescaling the tilting parameter $s$ and the SCGF $\lambda_{s,b}$ similarly to the previous Subsection, we obtain
\begin{equation}
    \label{eq:TwoStateSCGFLocal}
    \lambda_{\tilde{s},b} \approx - \log 2 + \frac{\tilde{\epsilon}_{\tilde{s}}}{2\sqrt{b}} \, ,
\end{equation}
with
\begin{equation}
    \label{eq:TwoStateEpsilon}
    \tilde{\epsilon}_{\tilde{s}} = \tilde{s} + \sqrt{4+\tilde{s}^2} \, .
\end{equation}
Eq.\ \eqref{eq:TwoStateSCGFLocal} describes fluctuations locally around $s_c$ for large (but finite) reward $b$. Also in this case, we plot in Fig.\ \ref{fig:BulkDanglingZoom} the function $\tilde{\epsilon}_{\tilde{s}}$ along with $b$-finite scalings.
\begin{figure}
    \centering
    \psfrag{a}{$\tilde{s}$}
    \psfrag{b}[][Bl][1][180]{$\tilde{\epsilon}_{\tilde{s}}$}
    \psfrag{c}{$b=15$}
    \psfrag{d}{$b=50$}
    \psfrag{e}{$b=250$}
    \psfrag{f}{$b \rightarrow \infty$}
    \includegraphics[width=0.5\linewidth]{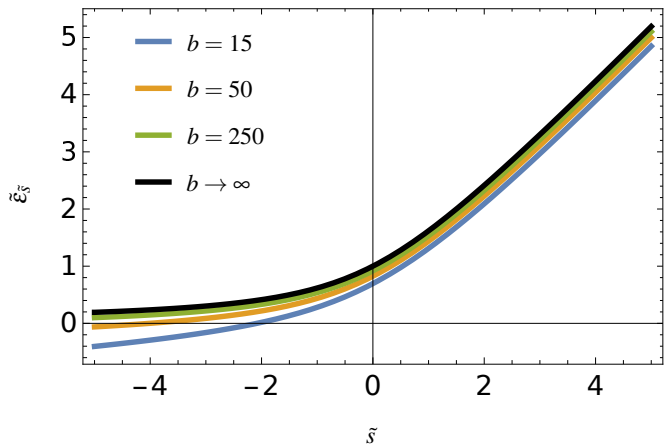}
    \caption{Crossover regime of the SCGF of the two-state Markov chain model around $s_c$ as a function of the scaling variable $\tilde{s}$ for different values of $b$. As $b$ increases, the colored curves collapse into the limiting curve predicted theoretically \eqref{eq:TwoStateEpsilon}. 
    }
    \label{fig:BulkDanglingZoom}
\end{figure}

The scaling of fluctuations is much different from what we found in \eqref{eq:BulkDanglingSCGFLocal} for the bulk-dangling model. We argue that the exact form of the scaling is not only determined by the dynamics of the model, but also by the topology of the graph considered. Indeed, the two-state Markov chain is only composed by two nodes playing the role of bulk and dangling chain, whereas in the bulk-dangling model, as previously mentioned, we count four key nodes (see Fig.\ \ref{fig:BulkDangling}) and among these, differently from the two-state Markov chain, the orange and red one play the role of a gate between the bulk and the yellow node of degree one.

To further corroborate this argument we also studied generalizations of the two-state Markov chain analyzed so far. These are obtained by considering as a probability to escape the rightmost state the value $b^{-\gamma}$, with $\gamma \geq 1$, and by rescaling, or not, the reward $b$ by $b^{-\gamma}$. In all cases considered (not shown here), the scaling of fluctuations around the critical point for the DPT are different from the case of the bulk-dangling model. These results support the argument that changing the dynamics does not make up for having different graph topologies. 

To investigate the robustness of the aforementioned DPT, we investigated also other variants of the two-state model, where the reward of the two nodes are set to $1$ and $k$ respectively. In this version of the model $k$ does not depend on $b$, and $1/b$ is only the transition probability to remain in state $k$. Interestingly, the DPT appears also in this case when $b$ goes to infinity, but both the critical tilting parameter $s_c$ and the behavior for $s<s_c$ depend on the value of $k$. This is in line with previous works on two state models \cite{Whitelam2018}, although with a remarkable difference. In these works, the author considers models where the transition matrix is symmetric, so that both nodes become absorbing in the appropriate limit. As a consequence, $s_c$ in those models is exactly $0$. In our work, instead, we are naturally led to consider non-symmetric transition matrices, such that only one of the two states becomes absorbing. To win this asymmetric absorbing dynamics, an infinitesimal $s$ is not sufficient, hence $s_c \neq 0$.

\section{Conclusions}
\label{sec:conclusion}

In this manuscript, we have shown the appearance of a first-order dynamical phase transition in two models that catch the relevant physical aspects of the dynamics of random walks on random graphs, for which the dynamical phase transition has hitherto only been argued. In both models, the random walk collects a cost---with general form given in \eqref{eq:Cost}---which scales differently in different regions of the graph. In the bulk-dangling model, very much similarly to a random walk on a random graph, the cost scales proportionally to the size of the graph in the bulk, whereas it gives only a constant contribution in the dangling chain. In the two-state Markov chain instead, we greatly simplified the topology of the graph and made the cost scale with a reward rather than keeping it linked to the graph structure. As a consequence, to keep the analogy with random walks on random graphs, we also suitably rescaled the transition probabilities inversely with the reward. We analyzed both models by applying a large deviation framework \cite{Carugno2022} that allowed us to carry out analytical results.

Remarkably, regardless of the precise details of the model, a first-order dynamical phase transition in the cost accumulated by the random walk always appears (see Fig.\ \ref{fig:BulkDangling} and \ref{fig:TwoState}). This is interpreted as a coexistence of paths that visit regions of the graph where the cost scales proportionally with the relevant physical parameter of the model (size $N$ or reward $b$) and paths that visit regions that only contribute to the cost with constant increments. We gave further evidence for this interpretation by also calculating the relevant driven process, which explains how fluctuations arise in time (see Eqs.\ \eqref{eq:BulkDanglingDriven1} and \eqref{eq:BulkDanglingDriven2}, and \eqref{eq:TwoStateDriven1} and \eqref{eq:TwoStateDriven2}). Furthermore, by zooming around the critical value for the transition, we exactly determined how fluctuations scale either with the system size $N$ or the reward $b$. Since the scaling turns out to be different in the two models investigated, we argue that although the dynamical phase transition is robust to topological changes in the model in the thermodynamic limit, the exact structure of the graph plays a role---along with the dynamics---for finite systems.

These results support the idea that also random walks on sparse random graphs undergo first-order phase transitions in the fluctuations of the mean-degree visited \cite{DeBacco2016,Coghi2019} for infinite-size graphs. However, a full proof has yet to be advanced. We believe that by implementing the large deviation framework discussed in \cite{Carugno2022} and in this manuscript one should be able to average the relevant action over the infinite realizations of the random graph ensemble, obtaining the scaled cumulant generating function in the thermodynamic-size limit. Furthermore, it would also be interesting to study transient regimes in time---and not only the asymptotics given by the large deviation theory---of the URW fluctuations. This could be done in principle by following ideas presented in \cite{Polettini2015, Causer2022, Carugno2022}.

Evidently, there is much scope for future work, both theoretical, as just mentioned, and applied. Related to the latter, for instance, by appropriately tuning the tilting parameter one could exploit the driven processes to generate optimal explorers of networks, a topic that has recently gained much attention \cite{Adam2019,Carletti2020}.

\bibliographystyle{ieeetr}
\bibliography{mybib}

\appendix
\section{Details on `Bulk-dangling model'}
\label{ref:app}

The exact form of the function $h$ appearing in \eqref{eq:BulkDanglingAction} reads
\begin{equation}
\label{eq:AppendixExacth}
\begin{split}
    h(s,N,\nu_{12}&,\nu_{23},\nu_{34},\nu_{44}) = \nu_{12} \left( \frac{3 s}{\bar{k}} - \log 2 \right) + \nu_{23} \left( \frac{N s}{\bar{k}} - \log (2 (N-2) ) \right) + \nu_{34}(N-3) \left( \frac{ (2N -5) s}{\bar{k}} - \log ((N-3)(N-2)) \right) + \\ 
    &+ \nu_{44}(N-3)(N-4) \left( \frac{(N-3) s}{\bar{k}} - \log(N_3) \right) + \nu_{12} \log \left( \frac{\nu_{12} + \nu_{23}}{\nu_{12}} \right) + \nu_{23} \left( \log \left( \frac{\nu_{12} + \nu_{23}}{\nu_{23}} \right) + \log \left( \frac{\nu_{23}+(N-3)\nu_{34}}{\nu_{23}} \right) \right)+ \\
    &+ \nu_{34} (N-3) \left( \log \left( \frac{\nu_{23}+(N-3)\nu_{34}}{\nu_{34}} \right)  + \log \left( \frac{\nu_{34} + (N-4) \nu_{44}}{\nu_{34}} \right) \right) + \nu_{44} (N-3)(N-4) \log \left( \frac{\nu_{34} + (N-4)\nu_{44}}{\nu_{44}} \right) + \\
    &+ \epsilon \left( 1 - 2\nu_{12} - 2\nu_{23} - 2(N-3) \nu_{34} - (N-3)(N-4) \nu_{44} \right) \, .
\end{split}
\end{equation}

The minimizers $\nu^*=(\nu^*_{12},\nu^*_{23},\nu^*_{34},\nu^*_{44})$ of the action \eqref{eq:BulkDanglingAction} are explicitly given by
\begin{align}
\begin{split}
\nu_{12}(s, N, \epsilon) &= \frac{a(s, N, \epsilon)}{(a(s, N, \epsilon) - 1)} (N - 3) \frac{b(s, N, \epsilon)}{(b(s, N, \epsilon) - 1)} \times \\ 
&\times \frac{((-1 + a(s, N, \epsilon)) (-1 + b(s, N, \epsilon)) (-1 + c(s, N, \epsilon) (-4 + N)))}{((-2 + a(s, N, \epsilon) (-1 + b(s, N, \epsilon)) (-2 + c(s, N, \epsilon) (-4 + N)) + (1 + b(s, N, \epsilon)) c(s, N, \epsilon) (-4 + N)) (-3 + N))}
\end{split}
\\[2ex]
\begin{split}
\nu_{23}(s, N, \epsilon) &= (N - 3) \frac{b(s, N, \epsilon)}{(b(s, N, \epsilon) - 1)} \times \\ 
&\times \frac{((-1 + a(s, N, \epsilon)) (-1 + b(s, N, \epsilon)) (-1 + c(s, N, \epsilon) (-4 + N)))}{((-2 + a(s, N, \epsilon) (-1 + b(s, N, \epsilon)) (-2 + c(s, N, \epsilon) (-4 + N)) + (1 + b(s, N, \epsilon)) c(s, N, \epsilon) (-4 + N)) (-3 + N))}
\end{split}
\\[2ex]
\begin{split}
\nu_{34}(s, N, \epsilon) &= \frac{((-1 + a(s, N, \epsilon)) (-1 + b(s, N, \epsilon)) (-1 + c(s, N, \epsilon) (-4 + N)))}{((-2 + a(s, N, \epsilon) (-1 + b(s, N, \epsilon)) (-2 + c(s, N, \epsilon) (-4 + N)) + (1 + b(s, N, \epsilon)) c(s, N, \epsilon) (-4 + N)) (-3 + N))}
\end{split}
\\[2ex]
\begin{split}
\nu_{44}(s, N, \epsilon) &= \frac{c(s, N, \epsilon)}{((N-4)c(s, N, \epsilon) - 1)} \times \\
&\times \frac{((-1 + a(s, N, \epsilon)) (-1 + b(s, N, \epsilon)) (-1 + c(s, N, \epsilon) (-4 + N)))}{((-2 + a(s, N, \epsilon) (-1 + b(s, N, \epsilon)) (-2 + c(s, N, \epsilon) (-4 + N)) + (1 + b(s, N, \epsilon)) c(s, N, \epsilon) (-4 + N)) (-3 + N))}
\end{split}
\end{align}

The inequality constraints that select the physical solution of \eqref{eq:BulkDanglingDeterminant} are
\begin{align}
\epsilon &> \frac{3 s}{2 \bar{k}} - \frac{\log 2}{2} \\
\epsilon &> - \frac{1}{2} \log \left( \frac{2(N-2)}{(N-2)e^{3 \frac{s}{\bar{k}}} + e^{N \frac{s}{\bar{k}}}} \right)\\
0 &> \tau^2 (N-3) +\tau (N-2)(N-4)e^{-(N-2) \frac{s}{\bar{k}}} \tau  - (N-2)(N-3)e^{-(2N-5) \frac{s}{\bar{k}}} \\
\epsilon &> (N-3) \frac{s}{\bar{k}} - \log(N-3) + \log(N-4) .
\end{align} 

\end{document}